\documentclass[aps, pre, preprint, groupedaddress, amsmath, amssymb]{revtex4-2}
\usepackage{graphicx}  
\usepackage{dcolumn}   
\usepackage{bm}        
\usepackage{multirow}
\begin{document}

\title{Texture Identification in Liquid Crystal-Protein Droplets using Evaporative Drying, Generalized Additive Modeling, and K-means Clustering }

\author{Anusuya Pal$^{a, b}$} \email{apal@g.ecc.u-tokyo.ac.jp}
\author{Amalesh Gope$^{c}$} 

\affiliation{$^{a}$ Department of Physics, Worcester Polytechnic Institute, MA, 01601, USA\\ $^{b}$ Graduate School of Arts and Sciences, The University of Tokyo, Komaba 4-6-1, Meguro, Tokyo, 153-8505, Japan \\ $^{c}$ Department of Linguistics and Language Technology, Tezpur University, Tezpur, Assam, 784028, India}



\begin{abstract}
Sessile drying droplets manifest distinct morphological patterns, encompassing diverse systems viz., DNA, proteins, blood, and protein-liquid crystal (LC) complexes. This study employs an integrated methodology that combines drying droplet, image texture analysis (features from First Order Statistics, Gray Level Co-occurrence Matrix, Gray Level Run Length Matrix, Gray Level Size Zone Matrix, and Gray Level Dependence Matrix), and statistical data analysis (Generalized Additive Modeling and K-means clustering). It provides a comprehensive qualitative and quantitative exploration by examining LC-protein droplets at varying initial phosphate buffered concentrations (0x, 0.25x, 0.5x, 0.75x, and 1x) during the drying process under optical microscopy with crossed polarizing configuration. Notably, it unveils distinct LC-protein textures across three drying stages: initial, middle, and final. The Generalized Additive Modeling (GAM) reveals that all the features significantly contribute to differentiating LC-protein droplets. Integrating the K-means clustering method with GAM analysis elucidates how textures evolve through the three drying stages compared to the entire drying process. Notably, the final drying stage stands out with well-defined, non-overlapping clusters, supporting the visual observations of unique LC textures. Furthermore, this paper contributes valuable insights, showcasing the efficacy of drying droplets as a rapid and straightforward tool for characterizing and classifying dynamic LC textures.

\end{abstract}


\maketitle


\section{Introduction}
\label{sec:intro}

Liquid crystals (LCs)-- a unique bridge between solid and liquid phases-- combine the flowing nature of liquids with the symmetrical properties of the crystals \cite{Pal2021}. These have optical anisotropy, exhibiting birefringence properties falling into two main categories-- thermotropic and lyotropic. The specific LC phase is determined by factors such as temperature changes (for thermotropic LCs) or variations in concentration (for lyotropic LCs). For instance, the thermotropic LC, 5CB transitions to a nematic phase at approximately $35^{\circ}$C \cite{pal2019phase}. In the nematic phase, the molecules are aligned in a specific direction, but lack a long-range positional order. When viewed under crossed polarizers, the nematic phase appears uniform and exhibits a characteristic ``worm-like" or ``thread-like" texture.  On the other hand, one typical example of a lyotropic LC is sodium dodecyl sulfate (SDS) in water. SDS is an amphiphilic molecule, meaning it has both a hydrophilic (water-attracting) head and a hydrophobic (water-repelling) tail. In certain concentrations and under specific conditions, SDS molecules can self-organize to form lyotropic LC phases. The lyotropic phases include lamellar (smectic), hexagonal, and cubic phases. Their textures exhibit distinctive patterns when observing these phases (lamellar, hexagonal, and cubic) under crossed-polarizing microscopy. It is important to note that these textures provide information about the molecular ordering within the LC phases. The specific patterns result from the molecular alignment and organization, which can be influenced by factors such as concentration, temperature, solvent, etc \cite{dierking2003textures}.

This emergence of the patterns has been investigated thoroughly from a soft matter perspective. Examples include studying copolymers, where different polymer blocks self-assemble \cite{yu2014photoresponsive}, exploring colloids that organize into structures like crystals and gels \cite{pawar2010fabrication}, understanding LC patterns with their distinctive molecular order \cite{de1993physics}, and investigating vortex formation in active matter, like swimming microorganisms displaying collective motion \cite{vicsek2012collective}. In contrast, patterns that emerge from drying sessile droplets, often found to have the ``coffee ring effect" \cite{deegan1997capillary}, have also been investigated in soft matter. When a liquid droplet containing solutes evaporates, it can leave behind distinctive patterns \cite{PAL2023102870}. Understanding the dynamics of a drying droplet and the resultant pattern formation is not only interesting from a fundamental physics perspective but also has practical applications in various fields, including inkjet printing \cite{he2017controlling}, coating technologies \cite{laborie2013coatings}, and bio-medical diagnostics \cite{PAL2023102870}. Recently, it has been investigated how the morphological patterns emerge when the optically active particles (5CB) are used as a probe in the different protein drying droplets \cite{pal2019phase, pal2019comparative, pal2022hierarchical}. It reveals that adding a fixed volume of LC to different globular protein solutions (lysozyme, BSA, and myoglobin added with de-ionized water) alters morphological patterns during drying \cite{anusuyabook2021}. In lightweight proteins (myoglobin and lysozyme), LCs are preferred to be randomly distributed. Conversely, in heavily weighted proteins like BSA, they form umbilical defect structures \cite{pal2019comparative}. Interestingly, when the solvent is replaced from the de-ionized water to buffer saline (PBS), the optical activities of the LCs become lower and lower as the initial PBS concentration increases from 0.25 to 1× \cite{pal2022hierarchical}.

Within the field of LCs, both traditional and deep learning approaches in the supervised machine learning (ML) have been applied during the last two decades. The traditional approaches include Random Forest (RF), Support Vector Machine (SVM), Decision Trees (DT), Multivariate Adaptive Regression, etc. In contrast, deep learning involves fast-forwarding  neural networks (NN) and artificial and convolution NN. SVM \cite{gong2008study} predicts the transition temperatures in thermotropic LC. Furthermore, RF \cite{chen2019random} and  ordinal networks \cite{pessa2022determining} are applied to forecast LC properties, while DT is specifically implemented to identify clearing temperatures in bent-core LCs \cite{antanasijevic2016prediction}. In contrast, the calibration of LC phases has been successfully predicted using neural networks (NN) \cite{grewal2006novel}. Even unsupervised ML is used to characterize the particle trajectories, pitch, and conical angle of the nematic phases relating to its structural and dynamical properties \cite{chiappini2020helicoidal}. A few recent investigations include predicting molecular ordering \cite{inokuchi2020predicting}, phase transitions \cite{osiecka2021investigation, sigaki2020learning, dierking2023deep}, topological defects \cite{walters2019machine, minor2020end}, and LC textures \cite{dierking2023testing, dierking2022classification}. On the other hand, different MLs are implemented for pattern recognition in a sessile drying droplet setting \cite{hamadeh2020machine, jeihanipour2022deep, harindran2022pattern}. 

Despite significant progress in machine learning (ML) and liquid crystals (LC), the integration of LC textures and ML techniques in a drying droplet scenario has been relatively limited. This paper addresses two main objectives. Firstly, we aim to quantify LC textures using advanced image processing techniques. Secondly, we employ generalized additive modeling and k-means clustering to identify distinct stages in the drying process. Specifically, we investigate which drying stage predominates in identifying different LC-textured droplets and determine the number of LC textures at each stage.

\begin{figure*}[h]
\centering
  \includegraphics[height=3cm]{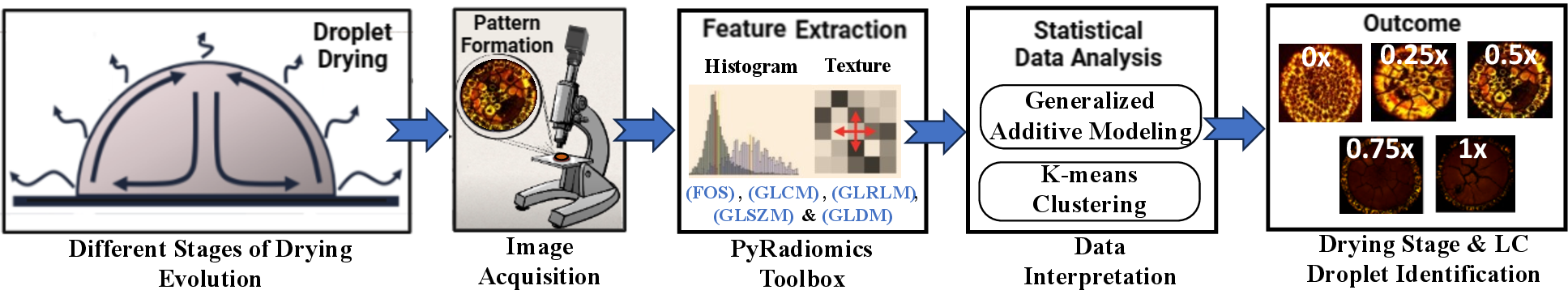}
  \caption{A flowchart showing the initiation of the drying evolution of a sessile droplet captured with optical microscopy. Different stages during the drying process are established. The quantitative image analysis using the \textit{Pyradiomics} toolbox is used to identify the dynamics of the liquid crystal (LC) textures induced by the drying process. The parameters include First Order Statistics (FOS), Gray Level Co-occurrence Matrix (GLCM), Gray Level Size Zone Matrix (GLSZM), Gray Level Run Length Matrix (GLRLM) and Gray Level Dependence Matrix (GLDM). These are feature vectors for Generalized Additive Modeling (GAM) and K-means Clustering to identify the LC-protein droplets at different initial buffered concentrations of 0x, 0.25x, 0.5x, 0.75x, and 1x.}
  \label{fig1}
\end{figure*}

To achieve this, we utilize texture analysis involving various order rank parameters, ranging from first-order to higher-order parameters. The quantitative image analysis employs the \textit{Pyradiomics} toolbox to unveil the dynamics of LC textures induced by the drying process. 

Figure~\ref{fig1} illustrates the flowchart outlining our approach. Each droplet progresses through various stages during the drying process, ultimately forming distinct fingerprint patterns. The drying droplets are examined using polarizing microscopy under crossed polarizing configurations. Throughout the drying process, features, including First Order Statistics (FOS), Gray Level Co-occurrence Matrix (GLCM), Gray Level Size Zone Matrix (GLSZM), Gray Level Run Length Matrix (GLRLM), and Gray Level Dependence Matrix (GLDM), are systematically extracted. Statistical data analysis is then applied to the diverse features extracted from the images. This step allows us to determine the number of LC textures present and identify which drying stage predominantly influences the identification of different LC-textured droplets. The comprehensive approach employed in this study aims to provide a detailed understanding of the evolving patterns in LC textures during the drying process.

\section{Methods}
\label{sec:exp}
\subsection{Samples and their preparations}

The lyophilized form of hen-egg white lysozyme (Catalog number L6876) was purchased from Sigma Aldrich, USA. The 1x PBS (phosphate buffer saline) is diluted to $0.75$, $0.5$, and $0.25$x. The $1$x PBS solution (Catalog number BP24384, Fisher BioReagents, USA) contains $0.137$M (${\sim}8.0$~mg mL$^{-1}$) NaCl, 0.002M (${\sim}0.2$~mg mL$^{-1}$) KCl, and $0.0119$M (${\sim}1.44$~mg mL$^{-1}$ of Na$_2$HPO$_4$ and ${\sim}0.24$~mg mL$^{-1}$ of KH$_2$PO$_4$) phosphates at a pH of ${\sim}7.4$. The $0$x represents the de-ionized water (Millipore, 18.2 M$\Omega$.cm at ${\sim}25^{\circ}$C). $100$~mg of lysozyme is weighed and mixed in $1$~mL of these PBS solutions. The thermotropic liquid crystal [5CB (4-Cyano-4'-pentylbiphenyl)] (Catalog number 328510, Sigma Aldrich, USA), was heated above its transition temperature (${\sim}35^{\circ}$C). ${\sim}10$~$\mu$L was added as a third component to the different protein-saline droplets. Therefore, we have five protein-LC droplets, with the initial PBS concentration of 0x, 0.25x, 0.5x, 0.75x, and 1x. 

A volume of ${\sim}1$~$\mu$L sample solution is pipetted on a freshly cleaned coverslip (Catalog number 48366-045, VWR, USA) under ambient conditions (the room temperature of ${\sim}25^{\circ}$C and the relative humidity of ${\sim}50$\%). The drying evolution is monitored every two seconds. The clock started when the droplets were deposited on the coverslips. The normalized time is calculated as the ratio of the instantaneous time to the total drying time. To ensure reliability, all these experiments were repeated three times. These samples show a good reproducibility. 

\subsection{Image acquisition}
The images were captured under $5$x magnification using cross-polarized optical microscopy (Leitz Wetzlar, Germany) configured in the transmission mode. An 8-bit digital camera (MU300, Amscope) was attached to the microscope, and the top-view images were clicked. 

\subsection{Image processing}
The \textit{Pyradiomics} tool \cite{van2017computational} in Python (version 3.10) extracts the quantitative features during drying. To do this, a mask is selected on the time series of the droplet to highlight the region of interest (ROI) for the feature extraction. This includes First Order Statistics (FOS), Gray Level Co-occurrence Matrix (GLCM), Gray Level Run Length Matrix (GLRLM), Gray Level Size Zone Matrix (GLSZM), and Gray Level Dependence Matrix (GLDM). The respective classes are \textit{RadiomicsFirstOrder}, \textit{RadiomicsGLCM}, \textit{RadiomicsGLRLM}, \textit{RadiomicsGLSZM}, and \textit{RadiomicsGLDM}. These classes compute the distribution of intensities within the specified ROI by initializing the batch of images as an input and the associated mask. 

For FOS, \(X\) is a set of \(N_p\) pixels in the ROI, \(P(i)\) is the first-order histogram with \(N_g\) discrete intensity levels, \(N_g\) is the number of non-zero bins, and \(p(i)\) is the normalized first-order histogram and equal to \(\frac{P(i)}{N_p}\). The FOS parameters include Mean ($\frac{1}{N_p} \sum_{i=1}^{N_p} X(i)$), Energy ($\sum_{i=1}^{N_p} (X(i) + c)^2$), RootMeanSquared ($\sqrt{\frac{1}{N_p} \sum_{i=1}^{N_p} (X(i) + c)^2}$); where c is used to shift the calculated intensities and prevent negative values in \(X\)), Entropy ($-\sum_{i=1}^{N_g} p(i) \log_2(p(i) + \epsilon)$, where \(\epsilon\) \(\approx 2.2 \times 10^{-16}\)), Skewness ($\frac{\mu_3}{\sigma^3} = \frac{1}{N_p} \sum_{i=1}^{N_p} \frac{(X(i) - \bar{X})^3}{\left(\frac{1}{N_p} \sum_{i=1}^{N_p} (X(i) - \bar{X})^2\right)^{3/2}}$), Kurtosis ($\frac{\mu_4}{\sigma^4} = \frac{1}{N_p} \sum_{i=1}^{N_p} \frac{(X(i) - \bar{X})^4}{\left(\frac{1}{N_p} \sum_{i=1}^{N_p} (X(i) - \bar{X})^2\right)^2}$, where \(\mu_3\), and \(\mu_4\) are the 3rd and 4th central moment, respectively), and Uniformity ($\sum_{i=1}^{N_g} p(i)^2$), and Variance ($\frac{1}{N_p} \sum_{i=1}^{N_p} (X(i) - \bar{X})^2$). 

A Gray Level Co-occurrence Matrix (GLCM) of dimensions \(N_g \times N_g\) characterizes the second-order joint probability function within the ROI image. It is defined as \(P(i, j|\delta, \theta)\), where \((i, j)\) represents the frequency of occurrences of levels \(i\) and \(j\) in pairs of pixels. These pixels are separated by a distance of \(\delta\) pixels along angle \(\theta\) from the center pixel. The feature value is computed individually for each angle in the GLCM, and the average of these values is obtained. Let \(P(i, j)\) denote the co-occurrence matrix for an arbitrary \(\delta\) and \(\theta\), and \(p(i, j)\) represent the normalized co-occurrence matrix, defined as \(P(i, j)/\sum P(i, j)\). \(N_g\) is the count of discrete intensity levels in the image, \(p_x(i) = \sum_{j=1}^{N_g} p(i, j)\) denotes the marginal row probabilities, and \(p_y(j) = \sum_{i=1}^{N_g} p(i, j)\) denotes the marginal column probabilities. Further, \(\mu_x\) is the mean gray level intensity of \(p_x\), calculated as \(\mu_x = \sum_{i=1}^{N_g} p_x(i) i\). Similarly, \(\mu_y\) is the mean gray level intensity of \(p_y\), defined as \(\mu_y = \sum_{j=1}^{N_g} p_y(j) j\). Additionally, \(\sigma_x\) represents the standard deviation of \(p_x\), and \(\sigma_y\) represents the standard deviation of \(p_y\).

The GLCM includes Contrast ($\sum_{i=1}^{N_g} \sum_{j=1}^{N_g} (i - j)^2 p(i, j)$), Difference Average ($DA$) ($\sum_{k=0}^{N_g-1} k p_{x-y}(k)$), Maximum Probability ($\max(p(i,j))$), Correlation ($\sum_{i=1}^{N_g} \sum_{j=1}^{N_g} p(i, j) \frac{ij - \mu_x \mu_y}{\sigma_x(i) \sigma_y(j)}$), Difference Entropy ($\sum_{k=0}^{N_g-1} p_{x-y}(k) \log_2(p_{x-y}(k) + \epsilon)$), Difference Variance ($\sum_{k=0}^{N_g-1} (k - DA)^2 p_{x-y}(k)$), Sum Entropy ($\sum_{k=2}^{2N_g} p_{x+y}(k) \log_2(p_{x+y}(k) + \epsilon)$), and Inverse Difference Moment (IDM) ($\sum_{k=0}^{N_g-1} \frac{p_{x-y}(k)}{1 + k^2}$).  

For GLSZM, it assesses the distribution of gray level zones in an image. A gray level zone is the count of connected pixels with the same gray level intensity. In the matrix \(P(i,j)\), the \((i,j)\)th element signifies the number of zones with gray level \(i\) and size \(j\) in the image. $N_g$ is the number of discrete intensity values in the image, $N_s$ is the number of discrete zone sizes in that image, $N_p$ is the number of pixels in the image, $N_z$ is the number of zones in the ROI, which is equal to $\sum_{i=1}^{N_g} \sum_{j=1}^{N_s} P(i,j) \text{ and } 1 \leq N_z \leq N_p$, and p(i,j) is the normalized size zone matrix, defined as $\frac{P(i,j)}{N_z}$. It is rotation-independent, and only one matrix is calculated for all directions within that ROI. 

The main parameters of GLSZM are Non-Uniformity ($\frac{\sum_{i=1}^{Ng} \left(\sum_{j=1}^{Ns} P(i,j)\right)^2}{N_z}$), Zone Non-Uniformity ($\sum_{j=1}^{Ns} \left( \sum_{i=1}^{Ng} P(i,j) \right)^2 / N_z$), Zone Entropy ($-\sum_{i=1}^{Ng} \sum_{j=1}^{Ns} p(i,j) \cdot \log_2(p(i,j) + \epsilon$), Zone Variance ($\sum_{i=1}^{Ng} \sum_{j=1}^{Ns} p(i,j) \cdot (j - \mu)^2$), and Variance ($\sum_{i=1}^{Ng} \sum_{j=1}^{Ns} p(i,j) (i - \mu)^2$, where \(\mu = \sum_{i=1}^{Ng} \sum_{j=1}^{Ns} p(i,j) i\)).

A Gray Level Run Length Matrix (GLRLM) characterizes sequences of pixels with the same gray level value, known as runs, by measuring their length in the number of consecutive pixels. The matrix, $P(i,j|\theta)$, represents the frequency of runs with gray level $i$ and length $j$ along the specified angle $\theta$ within the image ROI. $N_p$ is the number of pixels in the image, $N_r(\theta)$ is the number of runs in the image along angle $\theta$, which is defined as $\quad \sum_{i=1}^{N_g}\sum_{j=1}^{N_r} P(i,j|\theta) \quad \text{and } 1 \leq N_r(\theta) \leq N_p$. Therefore, similar to GLSZM, $P(i,j|\theta)$ is the run length matrix for an arbitrary direction $\theta$, and $p(i,j|\theta)$ is the normalized run length matrix, defined as $p(i,j|\theta) = \frac{P(i,j|\theta)}{N_r(\theta)}$. 

The GLRLM has four parameters. These are Gray Level Variance ($\sum_{i=1}^{N_g} \sum_{j=1}^{N_r} p(i,j|\theta) (i - \mu)^2$), Run Entropy ($-\sum_{i=1}^{N_g}\sum_{j=1}^{N_r} p(i,j|\theta)\log_2(p(i,j|\theta) + \epsilon)$), Run Variance ($\sum_{i=1}^{N_g}\sum_{j=1}^{N_r} p(i,j|\theta)(j - \mu)^2$, where $\mu = \sum_{i=1}^{N_g}\sum_{j=1}^{N_r} p(i,j|\theta)j$), and Gray Level Non-Uniformity ($\sum_{i=1}^{N_g} \left( \sum_{j=1}^{N_r} P(i,j|\theta) \right)^2 N_r(\theta)$).  

The Gray Level Dependence Matrix (GLDM) quantifies gray level dependencies in an image. Similar to GLSZM (which is dependent on the zone matrix), a gray level dependency is defined as the number of connected pixels within distance $\delta$ dependent on the center pixel. The neighboring pixel with gray level $j$ is dependent on the center pixel with gray level $i$ if $|i - j| \leq \alpha$. For GLDM, the parameters Dependence Entropy ($- \sum_{i=1}^{Ng} \sum_{j=1}^{N_d} p(i, j) \log_{2}(p(i, j) + \epsilon)$), Dependence Non-Uniformity ($\sum_{j=1}^{N_d} \left( \sum_{i=1}^{N_g} P(i, j) \right)^{2} N_z$), Dependence Variance ($\sum_{i=1}^{Ng} \sum_{j=1}^{Nd} p(i, j) (j - \mu)^{2}$, where $\mu = \sum_{i=1}^{N_g} \sum_{j=1}^{N_d} j p(i, j)$), Gray Level Non-Uniformity ($\sum_{i=1}^{N_g} \left( \sum_{j=1}^{N_d} P(i, j) \right)^{2} N_z$) and GrayLevelVariance ($\sum_{i=1}^{N_g} \sum_{j=1}^{N_d} p(i, j) (i - \mu)^{2}$, where $\quad \mu = \sum_{i=1}^{N_g} \sum_{j=1}^{N_d} i \cdot p(i, j)$). 

In total, there are thirty variables with eight First Order Statistics (FOS), eight Gray Level Co-occurrence Matrix (GLCM), four Gray Level Run Length Matrix (GLRLM), five Gray Level Size Zone Matrix (GLSZM), and five Gray Level Dependence Matrix (GLDM). 

\subsection{Statistical Data Analysis}

The generalized additive modeling (GAM) is implemented using R (version 4.1.2). For this, the \textit{library(mgcv)} and \textit{library(gam)} \cite{wood2001mgcv} were installed before the modeling. It involves selecting a basis for the space in which \( f \) resides. This selection leads to the basis functions \( F_j \) such that each is associated with parameters \( b_j\). The combination of these basis functions using these parameters results in $f(x) = \sum_{j=1}^{q} F_j(x) b_j$, where \( q \) represents the number of basis functions chosen for the representation of \( f(x) \).

The model is \textit{gam(LC-protein droplets $\sim$ s(Energy) + s(Entropy) + s(Kurtosis) + s(Mean Absolute Deviation) + s(Mean) + s(Robust Mean Absolute Deviation) + s(Root Mean Squared) + s(Skewness) + s(TotalEnergy) + s(Uniformity) + s(Variance) + s(Contrast) + s(Correlation) + s(Difference Average) + s(Difference Entropy) + s(Difference Variance) + s(Idm) + s(Maximum Probability) + s(Sum Entropy) + s(Gray Level Non-Uniformity) + s(Gray Level Variance) + s(Size Zone Non-Uniformity) + s(Zone Entropy) + s(Zone Variance) + s(Gray Level Non-Uniformity) + s(Gray Level Variance) + s(Run Entropy) + s(Run Variance) + s(Dependence Entropy) + s(Dependence Non Uniformity) + s(Dependence Variance) + s(Gray Level Non Uniformity) + s(Gray Level Variance), data = df)}, where the term ``LC-protein droplets" includes the initial PBS concentration of 0x, 0.25x, 0.5x, 0.75x, and 1x (five classes); and the term ``s" denotes the smooth function of the GAM modeling. The data is scaled using a \textit{scale} function before using the modeling to minimize the inter-scale differences. The summary of the model is described using \textit{summary(model)}. This modeling predicts five classes based on the smooth terms applied to the respective thirty predictor variables. 

Following the GAM modeling, K-means clustering is another complementary method used for data analysis. It does not involve using labeled training and test datasets like classification machine learning problems usually do. Instead, it focuses on discovering patterns, relationships, and structures (if there are any) within the dataset without predefined targets. K-means clustering \cite{sinaga2020unsupervised} is implemented in R using \textit{library(factoextra)} and \textit{library(cluster)}. It is built using \textit{kmeans(df, centers = 5, nstart = 25)}, where \textit{df} is the dataframe, \textit{centers} specify the number of clusters (or centroids) that the algorithm should aim to find in the data, and \textit{nstart} is the number of times the K-means algorithm should be run with different initial cluster centers. We chose to run the algorithm 25 times with different initializations, and the best result in minimizing the within-cluster sum of squares (WCSS) will be chosen. The clustering results are visualized using \textit{fviz\_cluster(km, data = df)}.

The comparative analysis considers all drying stages and the individual examination of three distinct drying stages. This approach aims to assess the impact of various stages on the characteristics of five protein-LC mixtures at the initial PBS concentrations of 0x, 0.25x, 0.5x, 0.75x, and 1x. GAM and K-means clustering facilitates a comprehensive understanding of the distinctive influences of each drying stage on these protein-LC droplets.

\section{Results and Discussions}
\label{sec:res}

\subsection{Qualitative and quantitative image analysis}
\label{subsec:image}

\begin{figure*}[h]
\centering
  \includegraphics[height=9.5cm]{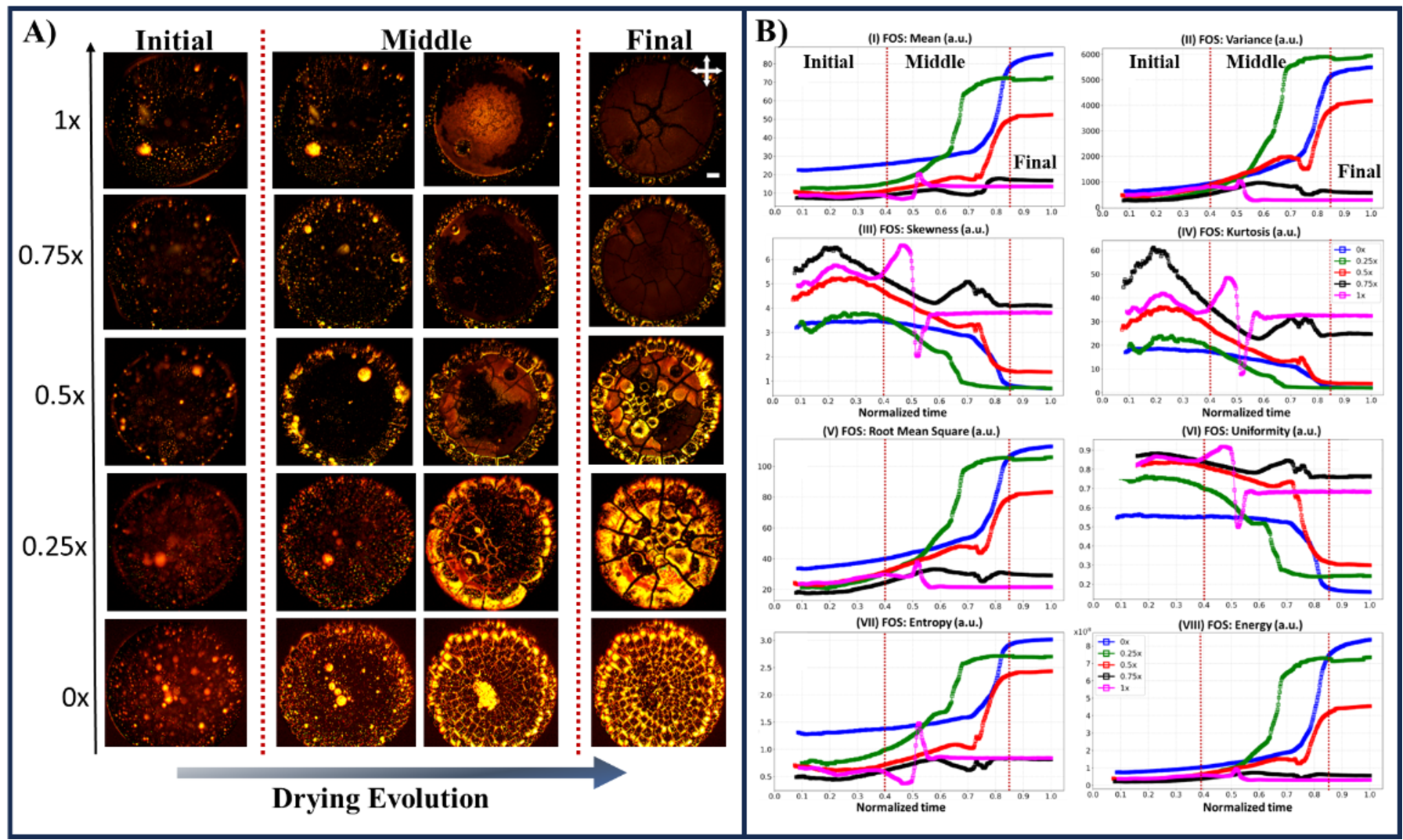}
  \caption{A) Optical images depict the progressive drying stages of liquid crystal (LC)-protein droplets with varied initial buffered concentrations (0x, 0.25x, 0.5x, 0.75x, and 1x) under a crossed polarizing configuration, represented by crossed arrows. The scale bar is 0.2 mm in length. B) First Order Statistics (FOS) encompass a range of variables:  (I) Mean, (II) Variance, (III) Skewness, (IV) Kurtosis, (V) Root mean square, (VI) Uniformity, (VII) Entropy, and (VIII) Energy. Dynamic changes in FOS parameters are presented over normalized time, calculated as the instantaneous time divided by the total time. Noteworthy stages-- initial, middle, and final--  are marked with red dashed lines. }
  \label{fig2}
\end{figure*}

Figure~\ref{fig2}A exhibits the qualitative drying evolution of the LC-protein droplets with varied initial PBS concentrations (0x, 0.25x, 0.5x, 0.75x, and 1x) under a crossed polarizing configuration. The 0x is the droplet of lysozyme, LC prepared in de-ionized water. During the drying process, the droplet height and contact angle decrease after pipetting the droplets onto the substrate. These droplets exhibit a spherical-cap shape. The curvature of these droplets induces higher mass loss near the periphery compared to the central region, leading to the well-known ``coffee-ring effect" \cite{deegan1997capillary}, commonly observed in various bio-colloids \cite{pal2020comparative}. The droplets become pinned to the substrate, and the lysozyme particles are transported through an outward capillary radial flow to compensate for the loss. The initial stage involves the general mechanism of the bio-colloidal droplets and does not depend on the varied concentrations of the LC-protein droplets. 

The middle drying stage is the most dynamic stage, which starts when the fluid front recedes from the periphery toward the central region and undergoes crack formation due to mechanical stress. The lysozyme droplet films are thin enough and buckled up as more water evaporates from the droplets. For 0x, upon mechanical stress-induced buckling of the protein-cracked domains, the LCs are drawn beneath these domains. The bright regions observed under crossed polarized configurations represent randomly oriented LCs. In contrast, the dark region in each domain corresponds to the attached protein layer that is not optically active, as discussed in detail in \cite{pal2019comparative}. When the initial PBS concentration goes from 0.25 to 1x, the interaction between lysozyme, salts, and LCs potentially influences the arrangement of lysozyme particles and LCs. However, LCs influence the packing and might not increase the film height. LCs appear trapped in the layer between lysozyme and salts in the central regions. The evaporation of a further volume of water leads to the bursting and random distribution of LCs. In contrast to 0x, an additional salt layer is on top of the LC distribution. This entire process offers a plausible explanation for the observed reduction in birefringence intensity under crossed polarizing configuration when PBS concentration increases from 0.25 to 1x, despite having a fixed volume of LCs. The detailed analysis can be found in \cite{pal2022hierarchical}. The final stage represents the concluding phase of the drying process, characterized by minor observable changes (see Figure~\ref{fig2}A). 

Figure~\ref{fig2}B(I-VIII) shows the quantitative drying evolution of the LC-protein droplets with varied initial PBS concentrations (0x, 0.25x, 0.5x, 0.75x, and 1x) under a crossed polarizing configuration using First Order Statistics (FOS). The variables include Mean, Variance, Skewness, Kurtosis, Root mean square, Uniformity, Entropy, and Energy. The three stages of the drying process exhibited the FOS values qualitatively. The initial stage of the drying process reveals gradual changes in these parameters. However, limited variations in these parameters are observed for specific PBS concentrations (0x, 0.25x, 0.5x, 0.75x, and 1x). The middle stage shows a rapid rise and is more vibrant than the initial and final stages. In contrast, the final stage stabilizes the parameters, where their values become relatively constant. This three-stage FOS evolution provides a comprehensive picture of the dynamic and settling behaviors exhibited by the droplet throughout drying. 

During the middle stage, a notable surge in the mean [Figure~\ref{fig2}B(I)], variance [Figure~\ref{fig2}B(II)], root mean square [Figure~\ref{fig2}B(V)], entropy [Figure~\ref{fig2}B(VII)], and energy [Figure~\ref{fig2}B(VIII)] is observed for PBS concentrations of 0x, 0.25x, and 0.5x. Conversely, a comparatively gradual increase is noted for concentrations of 0.75x and 1x during this stage. In contrast, uniformity [Figure~\ref{fig2}B(VI)] exhibits a declining trend in the middle stage compared to the initial stage. Uniformity, representing the homogeneity of the image array, is calculated as the sum of the squares of each intensity value.

Kurtosis, a measure of the ``peakedness" of the distribution values in the image ROI, indicates a higher concentration of mass towards the tails rather than the mean. Skewness measures the asymmetry of the distribution of values about the mean. In this study, skewness and kurtosis exhibit positive values; however, their magnitudes vary based on the initial PBS concentration and drying stage. Initially, these values are characterized by higher magnitudes, which diminish as the drying process progresses [Figure~\ref{fig2}B(III-IV)].

The mean represents the average gray level intensity within the ROI [Figure~\ref{fig2}B(I)]. At the same time, energy measures the magnitude of pixel values in the image [Figure~\ref{fig2}B(VIII)], with a more significant value indicating a greater sum of the squares of these values. Variance, as the mean of the squared distances of each intensity value from the mean, reflects the spread of the distribution about the mean [Figure~\ref{fig2}B(II)]. The root mean square (RMS) is derived from the square root of the mean of all squared intensity values, measuring the magnitude of the image values. The trend of Variance and RMS is the same [Figure~\ref{fig2}B(II and V)]. On the other hand, entropy quantifies the uncertainty and randomness in the image values, representing the average amount of information required to encode these values [Figure~\ref{fig2}B(VII)].

\begin{figure*}[h]
\centering
  \includegraphics[height=8cm]{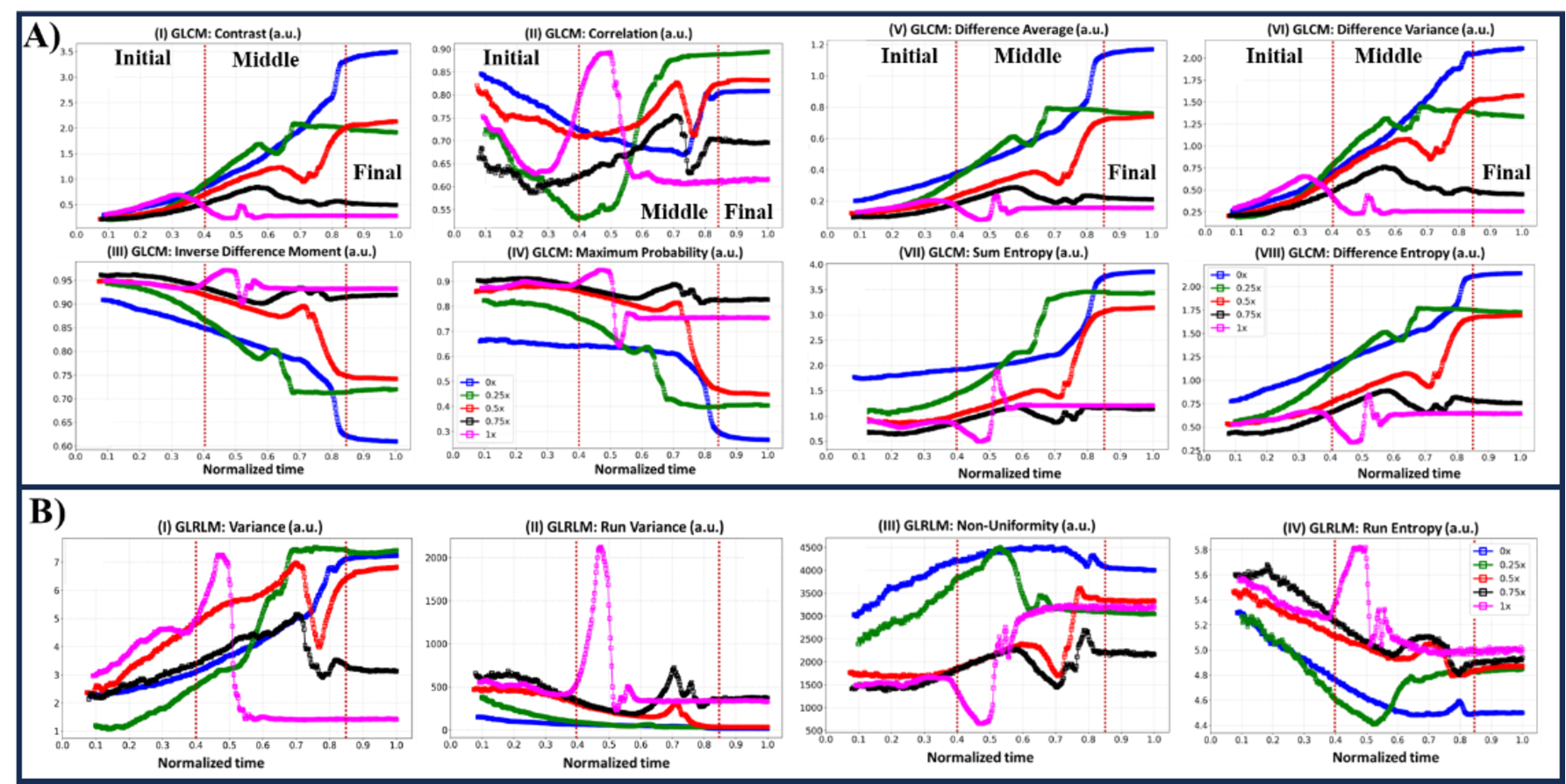}
  \caption{A)  Gray Level Co-occurrence Matrix (GLCM) encompass a range of variables: (I) Contrast, (II) Correlation, (III) Inverse Difference Moment, (IV) Maximum Probability, (V) Difference Average, (VI) Difference Variance, (VII) Sum Entropy, and (VIII) Difference Entropy. B) Gray Level Run Length Matrix (GLRLM) encompasses a range of variables: (I) Variance, (II) Run Variance, (III) Non-uniformity, and (IV) Run Entropy. Dynamic changes in GLCM and GLRLM parameters are presented over normalized time, calculated as the instantaneous time divided by the total time. The liquid crystal (LC)-protein drying droplets with varied initial buffered concentrations include 0x, 0.25x, 0.5x, 0.75x, and 1x. Noteworthy stages-- initial, middle, and final--  are marked with red dashed lines. }
  \label{fig3}
\end{figure*}

Figure~\ref{fig3}A(I-VIII) shows the quantitative drying evolution of the LC-protein droplets with varied initial PBS concentrations (0x, 0.25x, 0.5x, 0.75x, and 1x) under a crossed polarizing configuration using Gray Level Co-occurrence Matrix (GLCM). In the context of image analysis, various texture features derived from a GLCM provide valuable insights into local intensity patterns. The variables include Contrast, Correlation, Inverse Difference Moment (IDM), Maximum Probability, Difference Average, Difference Variance, Sum Entropy, and Difference Entropy [Figure~\ref{fig3}A(I-VIII)]. The middle stage is the most dynamic phase, characterized by peaks and troughs in the parameter values of GLCM. In contrast, a metric highlighting intensity variation tends to increase with more significant disparities among neighboring pixels [Figure~\ref{fig3}A(I)]. During the initial drying stage, the contrast among the LC-protein droplets remain primarily consistent, reflecting the overall uniformity in their initial flow behavior. However, a shift occurs as the drying process advances toward completion, and distinct droplets exhibit a more dominant contrast than others. This change in contrast dynamics signifies evolving interactions and variations in the drying patterns of individual droplets, adding complexity to the overall drying process.

On the other hand, Correlation measures the linear dependency of gray level values within the GLCM, where the value lies between 0 (uncorrelated) and 1 (perfectly correlated) [Figure~\ref{fig3}A(II)]. The current investigation reveals correlation values ranging from 0.55 to 0.99 through the drying process. Inverse Difference Moment (IDM) quantifies the local homogeneity of an image, with weights inversely proportional to Contrast weights [Figure~\ref{fig3}A(I and III)]. The Maximum Probability denotes the occurrences of the most predominant pair of neighboring intensity values [Figure~\ref{fig3}A(IV)]. The Difference Average reflects the relationship between occurrences of pairs with similar versus differing intensity values [Figure~\ref{fig3}A(V)]. Difference Variance emphasizes heterogeneity, assigning higher weights to differing intensity level pairs deviating more from the mean [Figure~\ref{fig3}A(VI)]. As observed, the heterogeneity values get higher at the later stage of drying. Sum Entropy captures the sum of neighborhood intensity value differences [Figure~\ref{fig3}A(VII)], while Difference Entropy gauges the randomness and variability in these differences [Figure~\ref{fig3}A(VIII)]. 

The Gray Level Run Length Matrix (GLRLM) computes Variance, Run Variance (RV), Non-Uniformity, and Run Entropy (RE) [Figure~\ref{fig3}B(I-IV)]. Variance quantifies the variation in gray level intensity for the runs, while RV assesses the variance in runs concerning run lengths. Specifically, RV is noticeably prominent during the middle drying stage when the PBS concentration is 1x, while variance is observed for all LC droplets throughout the entire drying process. The variance exhibits a trend starting with lower values and increasing towards the end of the process [Figure~\ref{fig3}B(I-II)]. Non-Uniformity quantifies the similarity of gray-level intensity values in the image, with a lower value indicating greater similarity in intensity values. On the other hand, RE measures the uncertainty and randomness in the distribution of run lengths and gray levels. A higher RE value suggests increased heterogeneity in the texture patterns. The expected relationship between Non-Uniformity and RE is that they should exhibit opposite trends, given that the former measures homogeneity, while the latter measures heterogeneity. This anticipated Contrast is observed in Figure~\ref{fig3}B(III-IV), where the initial drying stage is characterized by uniform texture, while the resulting patterns in the final drying stage are heterogeneous.

\begin{figure*}[h]
\centering
  \includegraphics[height=8cm]{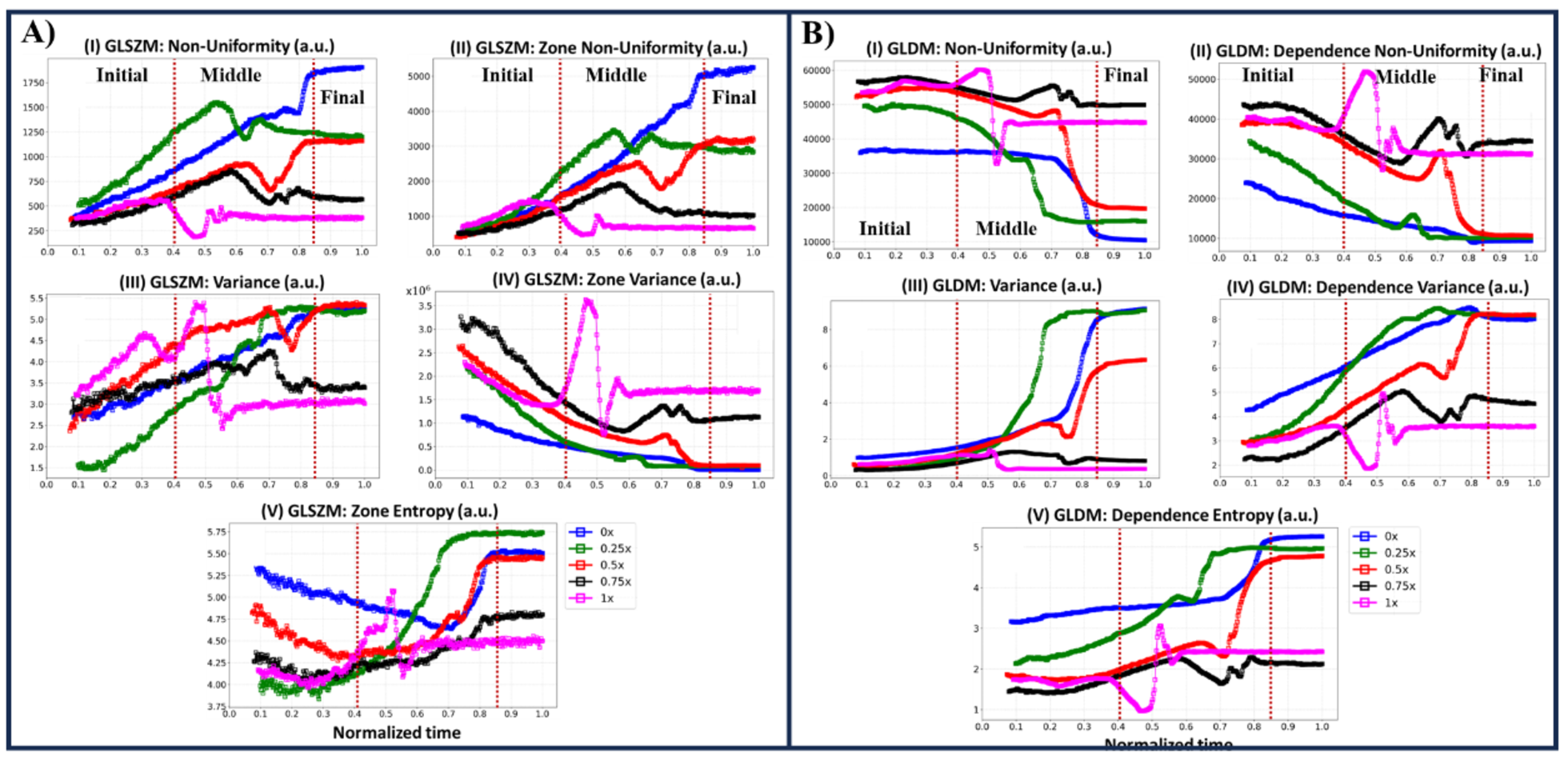}
  \caption{A)  Gray Level Size Zone Matrix (GLSZM) encompass a range of variables: (I) Non-Uniformity, (II) Zone Non-Uniformity, (III) Variance, (IV) Zone Variance, and (V) Zone Entropy. B)  Gray Level Dependence Matrix (GLDM)  encompass a range of variables: (I) Non-Uniformity, (II) Dependence Non-Uniformity, (III) Variance, (IV) Dependence Variance, and (V) Dependence Entropy. Dynamic changes in GLSZM and GLDM parameters are presented over normalized time, calculated as the instantaneous time divided by the total time. The liquid crystal (LC)-protein drying droplets with varied initial buffered concentrations include 0x, 0.25x, 0.5x, 0.75x, and 1x. Noteworthy stages-- initial, middle, and final--  are marked with the red dashed lines. }
  \label{fig4}
\end{figure*}

Figure~\ref{fig4}A-B(I-V) shows GLSZM features, including non-uniformity, zone non-uniformity, variance, zone variation (ZV), and zone entropy (ZE). In contrast, the GLDM feature consists of Non-Uniformity, Dependence Non-Uniformity(DN), Variance, Dependence Variance (DV), and Dependence Entropy. Variance quantifies the variance in gray level intensities for the zones, while ZV measures the variance in zone size volumes for the zones. ZE evaluates the uncertainty and randomness in the distribution of zone sizes and gray levels, with a higher value indicating increased heterogeneity in the texture patterns. Non-Uniformity assesses the variability of gray-level intensity values in the image, and a lower value signifies more homogeneity in intensity values. In contrast, Zone Non-Uniformity gauges the variability of size zone volumes in the image, with a lower value indicating more homogeneity in size zone volumes. 

The observed trend between Non-Uniformity and Zone Non-Uniformity follows a similar pattern, exhibiting lower values at the initial stage and higher values at the final drying stage. However, the trend in Variance and Zone Variance is not closely aligned, suggesting that the zone's impact may vary depending on the specific parameter being considered. Not only this, but the Gray Level Variance under the GLDM measures the variance in gray levels in the image, while DV assesses the variance in dependence size in the image. Though every parameter quantifies the variance, it is plausible that a specific parameter might be significantly higher for one protein-LC droplet while being negligible for others. The similar concept is also true for the ZE and DE. This variability underscores the intricate and context-dependent nature of the observed effects on different droplets [Figure~\ref{fig4}A-B(I-IV)]. 

Hence, qualitative and quantitative analyses affirm the predominant existence of three distinct stages throughout the drying process. This suggests that features derived from FOS, GLCM, GLRLM, GLSZM, and GLDM hold significant potential as key parameters for discerning and characterizing the LC-protein droplets.

\subsection{Quantitative statistical analysis: GAM and K-means clustering}
\label{subsec:stat}

GAM statistics show that all the features are equally crucial for identifying different LC-protein droplets. The predictors include all the feature parameters, whereas the response variable is the different LC-protein droplets. Table I-II describes the F-statistics, Degrees of freedom (Df), and p values obtained from GAM modeling. These are the different image texture features, such as FOS, GLCM, GLSZM, GLRLM, and GLDM. The smooth functions are applied to each feature in GAM modeling. The `s' in front of each feature indicates that these are smooth functions, for instance, s(Energy), s(Entropy), etc. Df represents the flexibility of the smooth functions, where the higher values indicate greater flexibility. Here, all parameters are equally flexible. The F-statistic measures the overall significance of the smooth term, and a higher value indicates a more significant effect. The p-values indicate whether the smooth term is statistically significant. The p-value less than 0.05 suggests a significant effect on the response variable (LC-protein droplets) due to the presence of all predictors (parameters within the features). This GAM output identifies the salient feature, and intriguingly, the results suggest that no single feature dominates; instead, all the features contribute significantly to the differentiation of LC-protein droplets. 

The output metrics from GAM show that the null deviance (a measure of the model's fit without any predictors) is 245.3621, with DF = 1941. When the predictors are included, the residual deviance drops to 1.8626, and DF = 1811, indicating the GAM model has significantly improved the fit compared to a null model. This underscores the intricate interplay of various image statistics in capturing the evolving textures during the drying process.

\begin{table}[ht]
\centering
\label{tab1}
\caption{Results of a Generalized Additive Model (GAM) with various features [First Order Statistics (FOS), and Gray Level Co-occurrence Matrix (GLCM)] and their corresponding parameters, degrees of freedom (Df), F-statistics, and p-values. }
\begin{tabular}{ccccc}
\text{Features} & \text{Parameters} & \text{Df} & \text{F-statistics} & \text{p values} \\
FOS & s(Energy) & 3 & 5.648 & 0.00075* \\
& s(Entropy) & 3 & 6.813 & 0.00014* \\
& s(Kurtosis) & 3 & 108.926 & $<$0.00001* \\
& s(Mean) & 3 & 5.527 & 0.00089* \\
& s(Root Mean Square) & 3 & 4.840 & 0.00233* \\
& s(Skewness) & 3 & 24.170 & $<$0.00001* \\
& s(Uniformity) & 3 & 5.483 & 0.00094* \\
& s(Variance) & 3 & 4.254 & 0.00528* \\
GLCM & s(Contrast) & 3 & 8.219 & $<$0.00001*  \\
& s(Correlation) & 3 & 75.518 & $<$0.00001*  \\
& s(Difference Average) & 3 & 7.759 & $<$0.00001*  \\
& s(Difference Entropy) & 3 & 7.036 & 0.00001* \\
& s(Difference Variance) & 3 & 8.705 & $<$0.00001* \\
& s(IDM) & 3 & 6.287 & 0.00030* \\
& s(Maximum Probability) & 3 & 3.446 & 0.01608* \\
& s(Sum Entropy) & 3 & 6.155 & 0.00036* \\
\end{tabular}
\end{table}

\begin{table}[ht]
\centering
\label{tab2}
\caption{Results of a Generalized Additive Model (GAM) with various features [Gray Level Run Length Matrix (GLRLM), Gray Level Size Zone Matrix (GLSZM), and Gray Level Dependence Matrix (GLDM)] and their corresponding parameters, degrees of freedom (Df), F-statistics, and p-values. }
\begin{tabular}{ccccc}
\text{Features} & \text{Parameters} & \text{Df} & \text{F-statistics} & \text{p values} \\
GLSZM & s( Gray Level Non-Uniformity) & 3 & 12.515 & $<$0.00001*  \\
& s(Gray Level Variance) & 3 & 86.396 & $<$0.00001*  \\
& s(Size Zone Non-Uniformity) & 3 & 4.135 & 0.00623* \\
& s(Zone Entropy) & 3 & 12.174 & $<$0.00001*  \\
& s(Zone Variance) & 3 & 3.892 & 0.00872* \\
GLRLM & s(Gray Level Non-Uniformity) & 3 & 24.496 & $<$0.00001*  \\
& s(Gray Level Variance) & 3 & 23.856 & $<$0.00001*  \\
& s(Run Entropy) & 3 & 54.181 & $<$0.00001*  \\
& s(Run Variance) & 3 & 5.054 & 0.00173* \\
GLDM & s(Dependence Entropy) & 3 & 8.728 & $<$0.00001*  \\
& s(Dependence Non-Uniformity) & 3 & 18.944 & $<$0.00001*  \\
& s(Dependence Variance) & 3 & 17.749 & $<$0.00001*  \\
& s(Gray Level Non-Uniformity) & 3 & 5.500 & 0.00092* \\
& s(Gray Level Variance) & 3 & 6.509 & 0.00022* \\
\end{tabular}
\end{table}

\begin{figure*}[h]
\centering
  \includegraphics[height=10cm]{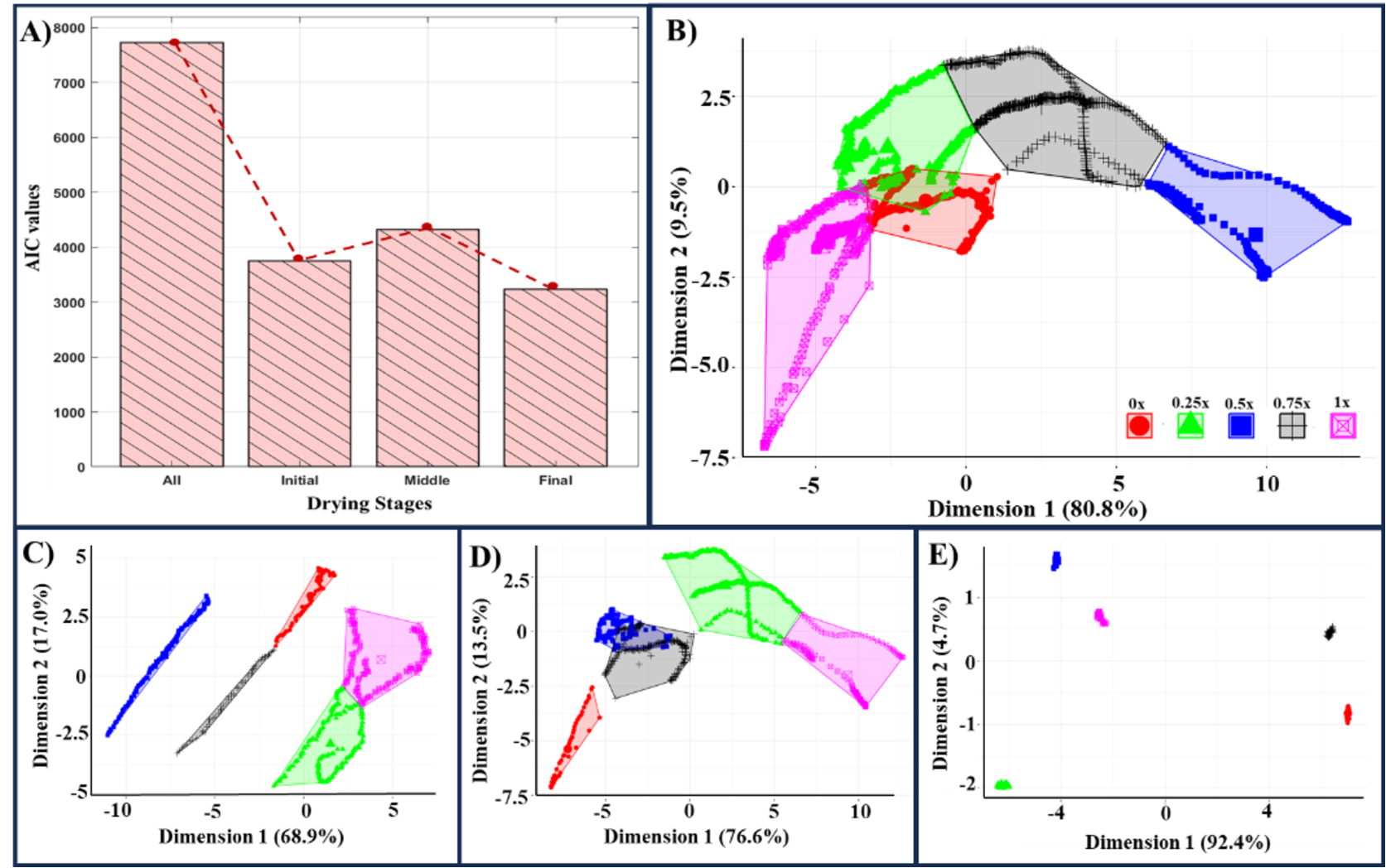}
  \caption{A) The Akaike Information Criterion (AIC) values of Generalized Additive Modeling (GAM) for different features for the liquid crystal (LC)-protein droplets at different drying stages-- the whole drying process (indicated with `all'), initial, middle, and final stages. The red dotted lines show the variation of the AIC values across these drying stages. The K-mean clustering where different initial buffered concentrations of 0x, 0.25x, 0.5x, 0.75x, and 1x are denoted by clusters for B) All, C) Initial stage, D) Middle stage, and E) Final stage of the drying process. }
  \label{fig5}
\end{figure*}

To find the drying stage for effectively distinguishing various LC-protein droplets, a Generalized Additive Model (GAM) analysis was conducted: initial, middle, final, and the entire drying process. The Akaike Information Criterion (AIC) compares the model's effectiveness in the drying stages. The negative AIC value provides a relative measure of 
the model's quality for the null model. Figure~\ref{fig5}A illustrates a comparative overview of Akaike Information Criterion (AIC) values across these stages, denoted as ``all" for the entire process. The AIC values obtained were 7720, 3749, 4325, and 3235 for the whole process, initial, middle, and final stages, respectively.

The AIC is a measure that balances the goodness of fit against the complexity of the model, with lower values indicating a superior trade-off. Intriguingly, the GAM results suggest that predicting different LC-protein droplets based on the drying process is not optimally achieved, as evidenced by the highest AIC value of 7720. Notably, the middle and initial stages follow in sequence, with AIC values of 4325 and 3749, respectively. Surprisingly, the final stage stands out, yielding the lowest AIC value of 3235, signifying a more favorable balance between model fit and simplicity. This implies that the final drying stage plays a pivotal role in effectively capturing and distinguishing the textures in various LC-protein droplets at the initial PBS concentrations of 0x, 0.25x, 0.5x, 0.75x, and 1x. 

In addition to the GAM analysis, we employed the K-means clustering method as a complementary approach to find the predominant drying stage influencing the identification of different LC-protein droplets [Figure~\ref{fig5}B-E]. This method involves grouping data points into clusters based on their similarities in a multidimensional space. Each plot's X and Y axes represent the first two principal components derived from the data containing all the predictors chosen for 2D visualization. These principal components are linear combinations of the original parameters, maximizing the variance in the data. Each point on the plot represents an observation, symbolized according to the initial buffered concentrations of 0x, 0.25x, 0.5x, 0.75x, and 1x.

When considering the entire drying process, the first two dimensions account for approximately $81\%$ and $10\%$ of the variance, respectively, explaining a total of around $91\%$ variance deviance. The concentration ellipses around each cluster reveal that 0.25x has a significant overlapping region with the 0x cluster and a smaller overlap with 1x. Additionally, the 0.5x cluster's peripheral boundary touches that of the 0.75x cluster [Figure~\ref{fig5}B], indicating that the five clusters are not uniquely determined by K-means clustering. The situation improves when focusing solely on the data from the initial drying stage [Figure~\ref{fig5}C]. Overlapping regions are minimal, with 0.5x being predominantly identified. In contrast, 0x and 0.75x have a touching boundary, similar to 0.25x and 1x. The first two dimensions represent approximately $69\%$ and $17\%$, explaining a total variance deviance of around $86\%$.

For the middle stage, the clusters of 0.5x and 0.75x exhibit overlapping regions with variance deviance of approximately $77\%$ and $14\%$, respectively [Figure~\ref{fig5}D]. Notably, K-means clustering provides distinct, spatially separable clusters without overlapping regions for the final drying stage. The two dimensions explain a total variance of approximately $97\%$ [Figure~\ref{fig5}E]. The shape and size of the clusters for each drying stage indicate how compact or dispersed the data points are within each cluster. Except for the final drying stage, the clusters are not well-separated and do not have similar sizes. For instance, in Figure~\ref{fig5}E, the clusters are distinct and balanced. The distribution and variation of data points within and across the clusters reflect how homogeneous these clusters become as the drying process progresses toward the end.

Therefore, the K-means clustering method and GAM analysis enhance our understanding of the evolving patterns in the drying process and provide valuable insights into the spatial distribution and homogeneity of different LC-protein droplets as the drying process progresses toward completion.

\section{Conclusions}
\label{sec:conc}
This paper presents a comprehensive qualitative and quantitative exploration of the dry- ing process, revealing distinct LC-protein textures across three main stages: initial, middle, and final. Through the application of image statistics, incorporating features from First Order Statistics (FOS), Gray Level Co-occurrence Matrix (GLCM), Gray Level Run Length Matrix (GLRLM), Gray Level Size Zone Matrix (GLSZM), and Gray Level Dependence Ma- trix (GLDM), the dynamics of these textures are analyzed concerning the drying stages. Generalized Additive Modeling (GAM) aims to identify salient features, and intriguingly, the results suggest that no single feature dominates; instead, all features contribute significantly to the differentiation of LC-protein droplets. This underscores the intricate interplay of various image statistics in capturing the evolving textures during the drying process. In addition, integrating the K-means clustering method with GAM analysis shows how the textures are influenced in the three drying stages. The final drying stage emerges with well-defined, non-overlapping clusters, supporting the visual observations and offering evidence of distinct LC textures at varying initial buffered concentrations (0x, 0.25x, 0.5x, 0.75x, and 1x). Therefore, this integrated approach enhances our understanding of the evolving morphological patterns and spatial distribution of LC-proteins in a sessile droplet configuration, contributing valuable insights into applying the drying droplet as a simple and rapid characterizing tool for identifying and classifying texture dynamics.

\section*{Author contribution}
Conceived and Designed: A.P., Data collection, and Compilation: A.P., Data Analysis and Interpretation: AP and AG, with A. G. supporting in statistical analyses, Conclusions: AP and AG, Initial Draft: AP, Revision, and Final editing: AG, and Overall Supervision: A.P.

\section*{Acknowledgments}

This research is supported by the Department of Physics at the Worcester Polytechnic Institute (WPI), USA, and the Graduate School of Arts and Sciences at The University of Tokyo, Japan. The authors would like to acknowledge Germano Iannacchione, Professor at WPI, USA, for the fruitful discussions. This research is partially supported by the Japan Society for Promotion of Science (JSPS), KAKENHI Grant No. 23KF0104. A. Pal expresses appreciation for the JSPS International Postdoctoral Fellowship for Research in Japan (Standard) for the period 2023-25.

\section*{Conflict of interest}
The authors declare no conflict of interest.

\section*{Data Availability Statement}
The data that support the findings of this study are available from the corresponding author upon reasonable request.

\bibliography{lcref}

\end{document}